\documentclass[prl,twocolumn,showpacs,floatfix]{revtex4}
\usepackage{amsfonts}
\usepackage{amssymb}
\usepackage{amsmath}
\usepackage{graphicx}

\newcommand{\beq}{\begin{equation}}
\newcommand{\eeq}{\end{equation}}

\newcommand{\bea}{\begin{eqnarray}}
\newcommand{\eea}{\end{eqnarray}}

\begin{document}

\title{How can we derive Fourier's Law from quantum mechanics? \\
 Exact master equation analysis} 

\author{Lian-Ao Wu and Dvira Segal}
\affiliation{
Chemical Physics Theory Group, Department of Chemistry, 
and Center for Quantum Information and Quantum Control, 
University of Toronto, \\
80 St. George Street, Toronto, Ontario M5S 3H6, Canada}

\begin{abstract}
We derive the macroscopic Fourier's Law of heat conduction from the
exact gain-loss time convolutionless quantum master equation under
three assumptions for the interaction kernel. To second order in the
interaction, we show that the first two assumptions are natural
results of the long time limit.
The third assumption can be satisfied by a family of interactions consisting an exchange effect.
The pure exchange model directly leads to energy diffusion in a weakly coupled spin-$1/2$ chain.
\end{abstract}

\maketitle

\textit{Introduction.---} Fourier's law, connecting the rate of heat
flow within a body to the temperature profile along the flow path,
is an empirical law based on observation. Despite its fundamental
nature, a derivation of this law from first principles is still
missing \cite{Bonetto00}.

In classical systems, extensive numerical simulations \cite{Lepri}
and rigorous derivations \cite{Bonetto00, Pereira, Closure}
manifested the applicability of  Fourier's law for specifically
designed nonlinear systems. It is still a great challenge to derive
this macroscopic law from microscopic {\it quantum} principles.
Quantum simulations of heat flow in spin chains pointed out on the
validity of Fourier's law, in connection with the onset of quantum
chaos \cite{Casati}. More recent studies have focused
on the derivation of this law from Schr\"odinger dynamics
\cite{Michel05, Michel06,Gemmer07}. By using the Hilbert space
average method, Michel \textit{et al.} demonstrated the emergence of
heat diffusion within a quasi degenerate level local Hamiltonian 
using the truncated Dyson series for the short-time displacement operator \cite{Michel05}.
Using this approach, the transition from diffusive to ballistic
dynamics at different length scales was further explored
\cite{Gemmer07}. These works successfully manifested the onset of
Fourier's law for a specific class of quantum models. The rigorous
derivation of this law from quantum dynamics for general 
Hamiltonians is still a great challenge.

\smallskip In this letter we employ the most general non perturbative 
microscopic master equation of the gain-loss form to derive quantum-mechanically the
Fourier's law. Unlike the derivation in Ref. \cite{Michel05},
applicable for a class of modular designed models, we use the most
general local Hamiltonian.
The diffusive behavior is obtained by making three
assumptions on the interacting kernel in the microscopic master
equation. These assumptions can be traced down to the structure of
coupling between local sites.
At weak coupling, going to the long time limit, 
we furthermore demonstrate that the first and second assumptions are 
independent of the specific form of the
interaction, and the exchange interaction plays the decisive role in
order for the third assumption to hold.

\textit{Master equation.--- } Consider a system with $N$ local units
(particles), each has the same $M$ (could be infinite) dimensional
eigenspace spanned by bases $\left\{ \left\vert n\right\rangle
\right\} $, possibly coupled to their own bathes. The eigenspace of
all $N$ particles is spanned by bases $\left\{ \left\vert
\mathbf{n}\right\rangle =\left\vert n_{1}\right\rangle
_{1}\left\vert n_{2}\right\rangle _{2}\left\vert n_{i}\right\rangle
_{i}...\left\vert n_{N}\right\rangle _{N}\right\}$. The most general
Hamiltonian of the system and its  bath is
\begin{equation}
H=H_{0}+\lambda H_{I},
\label{eq:Hamilton}
\end{equation}
where $H_{0}=H_{S}+H_{B}$ and $H_{S}$=$\sum_{i=1}^{N}
\sum_{n_i}\epsilon_{n_{i}}(i)\left\vert n_{i}\right\rangle
_{i}\left\langle n_{i}\right\vert $ is the system Hamiltonian. We
assume that the energy spectra of the sites are identical,
$\epsilon_{n}(i)=\epsilon_{n}(j)\geq 0$, and the ground state energy
is set as zero. $H_{B}$ is the bath Hamiltonian, consisting degrees
of freedom other than included in the system. The bath may act
locally on each site, $H_{B}=\sum_{i=1}^{N}H_{B}(i)$, where
$H_{B}(i)$ couples to the $^{i}$th local site. The second term
$H_{I}$ includes interactions between system particles, where
$\lambda$ characterizes the strength of these interactions. It can
be generally written as $H_{I}=\sum \left\vert
\mathbf{n}\right\rangle \left\langle \mathbf{m}\right\vert B_{
\mathbf{nm}}$, where $B_{\mathbf{nm}}$ 's are the matrix elements of
either system or bath operators. The dynamics  obeys the Liouville
equation $\frac{\partial }{\partial t}\,\rho _{tot}(t)=-i\lambda
\lbrack H_{I}(t),\rho _{tot}(t)]\equiv \lambda \mathcal{L}(t)\rho
_{tot}(t)$,$~$ where $\rho_{tot}$ is the density matrix of the total
system, and we work in the interaction representation
$H_{I}(t)=e^{iH_0t}H_Ie^{-iH_0t}$. The Liouville superoperator
$\mathcal{L}(t)$ is defined by this equation. The super projection
operation of interest, $\rho (t)=\mathcal{P}\rho _{tot}(t)$, defines
the \emph{relevant} part of the total density matrix for the open
system. This part exactly satisfies the time-local master equation
\cite{Breuer02}
\begin{equation}
\frac{\partial }{\partial t}\,\rho (t)=\mathcal{K}(t)\rho (t),
\label{eq:eq2}
\end{equation}
given that the initial state is in the relevant subspace,
$\mathcal{P}\rho_{tot}(0)=\rho_{tot}(0)$. The time-convolutionless
generator $\mathcal{K}(t)$ is in general an extremely complicated
object, calculated using perturbative expansions \cite{Breuer02}.
Though the time-local master equation (\ref{eq:eq2}) is less well
known than the Nakajima-Zwanzig equation
\cite{Nakajima,Zwanzig64}, it is easy to show that these forms are
equivalent \cite{Pereverzev06}. In order to project the diagonal
part of the total density $\rho _{tot}(t)$ we use the following
projection \cite{Pereverzev06,Zwanzig64}
\begin{equation}
\mathcal{P}\rho_{tot} =\sum \left\vert \mathbf{n}\right\rangle
\left\langle \mathbf{n}\right\vert \text{tr}\{\left\vert \mathbf{n}%
\right\rangle \left\langle \mathbf{n}\right\vert \rho
_{tot}\}\otimes \rho _{B}, \label{eq:eq200}
\end{equation}
where $\rho_{B}$ is the bath thermal equilibrium density matrix, and the trace
is taken over both system and bath states. Note that
$\rm{tr}\{\left\vert \mathbf{n}\right\rangle \left\langle
\mathbf{n}\right\vert \rho _{tot}\}=P_{\mathbf{n}}$ defines the
probability to find the system in state $\left\vert
\mathbf{n}\right\rangle $. The time-local master equation
 can be exactly written as \cite{note}
\begin{equation}
\frac{dP_{\mathbf{n}}}{dt}=\sum_{\mathbf{m}}W_{\mathbf{nm}}(t)P_{\mathbf{m}%
}-\sum_{\mathbf{m}}W_{\mathbf{mn}}(t)P_{\mathbf{n}},
\label{eq:eq20}
\end{equation}
where the rates, or interacting kernels $W_{\mathbf{nm}}(t)$, are
complicated functionals of the interaction $H_{I}(t)$. 
Assuming that $\mathcal{PL} (t)\mathcal{P}=0$, which is true
for closed systems \cite{Zwanzig64} and for many open systems \cite{Breuer02},
to second order in $H_I$ the expansion $\mathcal{K}
(t)=\int_{0}^{t}ds\mathcal{PL}(t)\mathcal{L}(s)\mathcal{P}$ holds.
One then obtains
\begin{equation}
W_{\mathbf{mn}}(t)=
2\lambda ^{2}{\rm Re}\int_{0}^{t}d\tau D_{\mathbf{nm}
}(\tau)e^{i(E_{\mathbf{n}}-E_{\mathbf{m}})\tau}
\label{eq:eq3}
\end{equation}
where
$B_{\mathbf{nm}}(\tau)=e^{iH_{B}\tau}B_{\mathbf{nm}}e^{-iH_{B}\tau}$,
and $D_{\mathbf{nm}}(\tau)=$
tr$_{B}\{B_{\mathbf{nm}}(\tau)B_{\mathbf{mn}}\rho _{B}\}$. The total
energy $E_{\mathbf{n}}= \sum_{i=1}^{N}\epsilon_{n_{i}}(i)$ is an
eigenvalue of $H_{S}$. For a closed system
$D_{\mathbf{nm}}(\tau)=\left\vert B_{\mathbf{mn} }\right\vert ^{2}$,
where $B_{\mathbf{nm}}$ 's are \textit{c}-numbers.

The gain-loss
master equation (\ref{eq:eq2}) was introduced for a \textit{closed} system, 
and proved vigorously in reference \cite{Zwanzig64}. In the long time limit, or
Markovian limit, the time-dependent interacting kernel $W(t)$
becomes a constant matrix for both open or closed systems
\cite{Lin74}. In what follows we focus on a closed system ($H_B$ is
neglected), and study energy diffusion between system units due to
the $H_{I}$ coupling. The starting point of our derivation is the exact
equation (\ref{eq:eq2}) with a {\it non-perturbative} kernel $W$.

\textit{Assumptions and derivation.--- } First consider a
one-dimensional system. We employ the nearest neighbor interaction
form $H_{I}=\sum_{i=1}^{N-1}V(i,i+1)$ \, where the symmetry
$V(i,i+1)=V(i+1,i)$ holds. Our {\it first assumption} is that the
interacting kernel matrix $W$ takes the same symmetry as the
interaction, i.e., $W=\sum_{i=1}^{N-1}W(i,i+1)$ and
$W(i,i+1)=W(i+1,i)$. This "localization" assumption implies that the
many-site correlation (kernel \textit{W}) is given by the sum of
two-site correlations $W(i,i+1)$. This assumption is not trivial as
the interacting kernels are not lineally related to the interaction
$H_{I}$. The matrix elements of $W$ are therefore given by
\begin{equation}
W_{\mathbf{nm}}(t)=\sum_{i=1}^{N-1}W_{n_{i}n_{i+1}},_{m_{i}m_{i+1}}(i,i+1;t)%
\prod_{j\neq i,i+1}\delta _{n_{j}m_{j}}.  \label{eq:eq50}
\end{equation}
As diagonal elements ($\mathbf{n}=\mathbf{m}$) do not contribute to
equation (\ref{eq:eq20}), they are allowed to be exceptions of the
assumption. Our {\it second assumption} describes energy
conservation between initial and final system states,
\bea
&&W_{n_{i}n_{i+1}},_{m_{i}m_{i+1}}(i,i+1;t) =
\label{eq:eq5}
\nonumber\\
&&\bigg\{
\begin{array}{c}
w_{n_{i}n_{i+1}}(i,i+1;t);\text{  } n_{i}=m_{i+1}\text{ and } n_{i+1}=m_{i} \\
0;\text{  Otherwise.}
\label{eq:second}
\end{array}
\eea
Besides energy conservation, this condition also implies that the
local spectra are {\it anharmonic}, see discussion after Eq.
(\ref{eq:W}). In the\emph{\ third assumption}, we assume that
$w_{n_{i}n_{i+1}}(i,i+1;t)=w(i,t)$, independent of the 
$n_{i}$ and $n_{i+1}$ quantum numbers. 
The probability to find the $^{i}$th particle in state $\left\vert
n_{i}\right\rangle _{i}$ is
\begin{equation}
P_{n_{i}}(i)=\text{tr}\{\left\vert n_{i}\right\rangle _{i}\left\langle
n_{i}\right\vert \rho _{tot}\}=\sum_{n_{j}\neq n_{i}}P_{\mathbf{n}}.
\label{eq:eq6}
\end{equation}
Incorporating Eqs. (\ref{eq:eq50}), (\ref{eq:eq5}) and
(\ref{eq:eq6}) into (\ref{eq:eq20}) we obtain
\begin{eqnarray}
\frac{dP_{n}(i)}{dt} &=&w(i,t)[P_{n}(i+1)-P_{n}(i)]-
\label{eq:rate}
\nonumber\\
&&w(i-1,t)[P_{n}(i)-P_{n}(i-1)],
\end{eqnarray}
where for convenience we use the short notation $P_n(i)$. In
deriving (\ref{eq:rate}) we have also utilized the symmetry
$W(i,i+1)=W(i+1,i)$ which holds exactly for a closed system at second
order  [see text after Eq. (\ref{eq:eq3})]. Next we write an
equation of motion for the internal energy at each site, $u(i)=\sum
\epsilon_{n}P_{n}(i)$,
\begin{eqnarray}
\frac{du(i)}{dt} &=&w(i,t)(u(i+1)-u(i))
\label{eq:eq8}
\nonumber\\
&&-w(i-1,t)(u(i)-u(i-1)).
\end{eqnarray}
The continuous version of this equation is
\begin{equation}
\frac{\partial u(x,t)}{\partial t}=a^{2}\frac{\partial }{\partial
x}\left( w(x,t)\frac{\partial u(x,t)}{\partial x}\right),
\label{eq:cont}
\end{equation}
where $a$ is the distance between neighboring sites. Generalization
of this derivation to a three dimensional simple cubic lattice is
straightforward, leading to Eq. ({\ref{eq:cont}}) with $\mathbf{r}$
and $\mathbf{\nabla}$ replacing $\frac{\partial}{\partial x}$ and
$x$. Applying the continuity equation for the energy density,
$\frac{\partial u(\mathbf{r},t)}{\partial t}=-%
\mathbf{\nabla }\cdot \mathbf{J}(\mathbf{r},t)$, $\mathbf{J}$ is the heat current, 
we exactly obtain the Fourier's law 
\bea
\mathbf{J}(\mathbf{r},t)=-w(\mathbf{r},t)a^{2}\mathbf{\nabla }u(
\mathbf{r},t)=-\kappa (\mathbf{r},t)\mathbf{\nabla
}T(\mathbf{r},t). 
\label{eq:Fourier}
\eea
Here $T(\mathbf{r},t)$ denotes the temperature
profile and $\kappa (\mathbf{r},t)=w(\mathbf{r},t)a^{2}C_{V}$ with
$C_{V}=\partial u/\partial T$ as the specific heat.
The heat conductivity $\kappa$ is essentially time-dependent due to the explicit time dependence of
the microscopic rates $w({\mathbf r},t)$. 
In the long time, or Markovian limit, the microscopic rates and the temperature profile become 
constants, 
leading to a time-independent relation.
Equation (\ref{eq:Fourier}) is the main result of our paper.
We emphasize that it was derived from the exact
master equation (\ref{eq:eq2}) for a generic local Hamiltonian.

\textit{An exact model in the second order.---} We present next a
model Hamiltonian that exactly satisfies the three assumption
leading to Fourier's law in the Markovian limit. We assume a closed
system ($H_B$ is neglected), and employ a pure exchange interaction
form
\begin{equation}
H_{I}=\sum_{i=1}^{N-1}J_{i,i+1}\mathcal{E}^{(i,i+1)}.
\label{eq:eq9}
\end{equation}
Here $\mathcal{E}^{(i,i+1)}$ is the permutation operator and
$J_{i,i+1}$ are nearest-neighbor coupling constants (superexchange
for spin system), taken as constants $J_{i,i+1}=J$ and set to one in
the following discussion. In the two-level case ($M=2$), since
$\mathcal{E}^{(i,i+1)}=\frac{1}{2}(1+\mathbf{\sigma} _{i}\cdot
\mathbf{\sigma }_{i+1})$, Eq. (\ref{eq:eq9}) is simply the
Heisenberg spin-$1/2$ exchange interaction model. 
The correlation function in (\ref{eq:eq3}) is given by
\begin{equation}
D_{\mathbf{nm}}(\tau)=\sum_{i=1}^{N-1}\delta _{n_{i}m_{i+1}}\delta
_{n_{i+1}m_{i}}\prod_{j\neq i,i+1}\delta _{n_{j}m_{j}},
\end{equation}
when $\mathbf{n\neq m}$, satisfying the first assumption.
Furthermore, the second assumption is fulfilled here exactly as the
matrix $W$ has the exact form of Eq. (\ref{eq:eq5}) with
$w(i,t)=2\lambda^{2}t$. In order to check the validity of the second
order approximation, we compare this result with the exact solution
for a three unit systems, $N=3$ and $M=2$. Let the initial state be
$\left\vert 1\right\rangle _{1}\left\vert 0\right\rangle
_{2}\left\vert 0\right\rangle _{3}$. We can get an exact solution
for the dynamics for the interaction (\ref{eq:eq9}),
$dP_{1}(2)/dt=w(t)(P_{1}(1)+P_{1}(3)-2P_{1}(2))$, where
$w(t)=\frac{\sqrt{2} \lambda \sin (2\sqrt{2}\lambda t)}{2\cos
^{2}(\sqrt{2}\lambda t)-1}$. For weak coupling (small $\lambda$),
$w(t)=2\lambda ^{2}t$, which is the same as the second order result.
While this model satisfies all three assumptions, in the long time
limit it does not lead to Fourier's law.

However, the energy difference between spin states of site $i$ may
slightly differ from that of site $i+1$, $\Delta
_{n_{i},n_{i+1}}(i)=[\epsilon_{n_{i}}(i)-\epsilon_{n_{i+1}}(i)]-
[\epsilon_{n_{i}}(i+1)-\epsilon_{n_{i+1}}(i+1)] \neq 0$. This effect
may originate from thermal fluctuations due to the existence of
local bathes at each site, which do not contribute to the diffusive
behavior. In this case, $w_{n_{i}n_{i+1}}(i,t)=2\lambda
^{2}\frac{\sin (\Delta _{n_{i},n_{i+1}}(i))t)}{\Delta
_{n_{i},n_{i+1}}(i)}$. At short times, if the difference
$\Delta_{n_{i},n_{i+1}}(i)$ is small, one again finds $w(t)\propto
t$. In contrast, at long times $w_{n_{i}n_{i+1}}(i)\rightarrow
2\pi \lambda ^{2}\delta (\Delta _{n_{i},n_{i+1}}(i))$.
When the energy spectrum of each spin state forms a band that is
dense enough \cite{GemmerPhysicaE05}, the Fermi's Golden rule is
obtained, $w_{n_{i}n_{i+1}}(i)=2\pi\lambda ^{2}\Gamma (0)$, where
$\Gamma(0)$ is the density of states at zero detuning.

Thus, this simple exchange model exactly reproduces the long time Fourier's
law (\ref{eq:Fourier}) with $\kappa=2 \pi \lambda^2 a^2 C_V \Gamma(0)$.
This result manifests that the local sites do not need to acquire
exactly identical spectra. 
It also provides a microscopic explanation of the validity of
Fourier's law in spin chains with Heisenberg-type interactions
\cite{Sologubenko01}.


\textit{Long Time (Markovian) Limit.---} 
Next we show that at second order in the interacting kernel
the first two assumptions [Eqs. (\ref{eq:eq50})-(\ref{eq:second})]
are model-independent in the long time limit, and the third
assumption is valid for family of interactions with exchange effect.
We consider the most general interaction $V(i,i+1)$ for a one-dimensional system. 
It is easy to show that
\beq
B_{\mathbf{nm}}=\sum_{i=1}^{N-1}V_{n_{i}n_{i+1}},_{m_{i}m_{i+1}}(i,i+1)
\prod_{l\neq i,i+1}\delta _{n_{l}m_{l}}. \eeq
In the long time (Markovian) limit Eq. (\ref{eq:eq3}) reduces to
\bea
W_{\mathbf{mn}}(t\rightarrow \infty )=2\pi\lambda^{2}\left\vert B_{\mathbf{mn}%
}\right\vert ^{2}\delta (E_{\mathbf{n}}-E_{\mathbf{m}}), \eea
using $\frac{\sin ((E_{\mathbf{n}}-E_{\mathbf{m}})t)}{(E_{\mathbf{n}}-E_{%
\mathbf{m}})}\rightarrow \pi \delta (E_{\mathbf{n}}-E_{\mathbf{m}})$ when $%
t\rightarrow \infty$. Assuming the bound states are non-degenerate,
a somewhat tedious calculation shows that the first assumption is
satisfied, and the off-diagonal matrix elements are given by
\begin{eqnarray}
& & W_{n_{i}n_{i+1}},_{m_{i}m_{i+1}}(i,i+1;t)=
\nonumber\\
& &2\pi\lambda ^{2}\left\vert
V_{n_{i}n_{i+1}},_{m_{i}m_{i+1}}(i,i+1)\right\vert ^{2}\delta
(\Delta _{_{n_{i}n_{i+1}},_{m_{i}m_{i+1}}}),
\nonumber\\
\label{eq:W}
\end{eqnarray}
where $\Delta
_{_{n_{i}n_{i+1}},_{m_{i}m_{i+1}}}=\epsilon_{n_{i}}+\epsilon_{n_{i+1}}-\epsilon_{m_{i}}-\epsilon_{m_{i+1}}$.
Since $\epsilon_{n}\geq 0$, the delta function in Eq. (\ref{eq:W})
implies that  $\epsilon_{n_{i}}=\epsilon_{m_{i+1}}\ $ and
$\epsilon_{n_{i+1}}=\epsilon_{m_{i}}$, provided that the energy
spectra are {\it anharmonic}. Therefore, in the non-degenerate case,
only  transitions between the quantum states $n_{i}=m_{i+1}$ and
$n_{i+1}=m_{i}$ exist, in accordance with the second assumption. In
contrast, when the energy spectra are {\it strictly harmonic}, the
delta function can be satisfied for large number of combinations,
$\epsilon_{n_i}-\epsilon_{m_{i}}=\epsilon_{m_{i+1}}-\epsilon_{n_{i+1}}=j
\omega_0$. Here $\omega_0$ is the energy difference between subunit
states (equal for all $N$) and $j$ is an integer. In this case the
second assumption does not hold, and we cannot derive Eq.
(\ref{eq:rate}) and the subsequent result (\ref{eq:Fourier}). Thus,
interestingly, in order to derive Fourier's law,
the system spectrum should be anharmonic, in
accordance with classical results \cite{Bonetto00,Lepri}.

At second order, the first two assumptions are therefore the results of the Markovian
limit. As discussed above, we can expect that there is a dense band
structure around each level $\epsilon_{n}$, originating from bath
fluctuations or quantum tunneling between sites. Defining the
density of states at the energy difference around $\Delta $ as
$\Gamma (\Delta )$, we obtain \cite{GemmerPhysicaE05}
\begin{equation}
w_{n_{i}n_{i+1}}(i)=2\pi\lambda ^{2}\Gamma (0)\left\vert
V_{n_{i}n_{i+1}},_{n_{i+1}n_{i}}(i,i+1)\right\vert ^{2},
\end{equation}
which manifests that the third assumption is valid, depending on the form of
specific Hamiltonians. We calculate next the microscopic rates for different types of
interactions. Assuming translational symmetry in the system,
we need only discuss a pair of sites, for instance between site 1 and 2,
$V_{n_{1}n_{2}},_{n_{2}n_{1}}(1,2)=\left\langle n_{1}n_{2}\right\vert
V(1,2)\left\vert n_{2}n_{1}\right\rangle$.

In general, particles at each site may be electrons or atoms. To
simplify, we consider the case with one particle at each site. Our
first example is the short-range delta interaction, $V(1,2)=g\delta
(r_{12})$, where $r_{12}=\left\vert x_{2}-x_{1}\right\vert$, and
$x_{1}(x_{2})$ is the coordinate of the first (second) particle.
Physically, this interaction describes particles that move almost
independently in the site interior, while collisions, leading to
energy exchange between particles, occur at the edge points
$x_{1}=x_{2}$. This picture is appropriate for describing phonon
collisions in solids \cite{Peierls01}. For this type of interaction
the matrix elements become
\begin{equation}
V_{n_{1}n_{2}},_{n_{2}n_{1}}(1,2)=g\int dx\left\vert \phi
_{n_{1}}(x)\right\vert ^{2}\left\vert \phi _{n_{2}}(x)\right\vert
^{2}. \label{eq:short}
\end{equation}
This integral is almost independent of quantum numbers $n_{1}$
and $n_{2}$ for many systems \cite{Noya59}. Therefore, the third
assumption is generally valid for short-range interactions. For
example, in one-dimensional infinite square well (width $d$) with
the wave function $\phi _{n}(x)=\sqrt{\frac{2}{d}}\sin \frac{n\pi
x}{d}$, the matrix element $V_{n_{1}n_{2}},_{n_{2}n_{1}}(1,2)=1/d$
is completely independent of quantum numbers, assuming $d>a$ so that
two particles could collide.

Our second example is the long-range interaction.
We choose the general form
$V(1,2)=V(r_{12})(\lambda_1+\lambda_2\mathcal{E} ^{(1,2)})$, where
the $\lambda$'s are constants and $\mathcal{E}^{(1,2)}$ is the
spacious exchange operator \cite{Fetter}. Considering the first
order contribution to this potential, $V(1,2)=\left\vert
x_{2}-x_{1}\right\vert (\lambda_1+\lambda_2\mathcal{E}^{(1,2)})$,
the matrix elements $V_{n_{1}n_{2}},_{n_{2}n_{1}}(1,2)$ become the
sum of the direct ($D$) and  exchange ($E$) terms,
\bea
&&V_{n_{1}n_{2}}^{D},_{n_{2}n_{1}}(1,2) =
\nonumber\\
&&\lambda_1\int dx_{1}dx_{2}\phi _{n_{1}}^{\ast }(x_{1})\phi _{n_{2}}^{\ast
}(x_{2})\left\vert x_{2}-x_{1}\right\vert \phi _{n_{1}}(x_{2})\phi
_{n_{2}}(x_{1}),
\nonumber\\
&&V_{n_{1}n_{2}}^{E},_{n_{2}n_{1}}(1,2) =
\nonumber\\
&&\lambda_2\int dx_{1}dx_{2}\left\vert \phi _{n_{1}}(x_{1})\right\vert
^{2}\left\vert \phi _{n_{2}}(x_{2})\right\vert ^{2}\left\vert
x_{2}-x_{1}\right\vert.
\eea
In what follows we take $\lambda_1=\lambda_2=1$.
If the overlap between the wave functions of two particles is zero,
the direct integral diminishes because of the orthogonality of two
states, and the exchange integral ($x_2>x_1$) is
$V_{n_{1}n_{2}}^{E},_{n_{2}n_{1}}(1,2)=a+\left\langle
n_{2}\right\vert x_{2}^{\prime }\left\vert n_{2}\right\rangle
-\left\langle n_{1}\right\vert x_{1}\left\vert n_{1}\right\rangle $,
where $x_{2}^{\prime }=x_{2}-a$ is the relative coordinate of
particle 2.
If the wavefunction  $\phi _{n}(x)$ has a well defined parity,
$\left\langle n\right\vert x\left\vert n\right\rangle =0$ for the
relative coordinate of each particle, implying that
$V_{n_{1}n_{2}},_{n_{2}n_{1}}(1,2)=a$ is a constant. This satisfies
the third assumption. Note that the result is independent of the
details of the local Hamiltonian.

If the overlap is not zero, yet small, the direct integral is
expected to be small. Considering again the one-dimensional infinite
square well, when $a/d<1$, two particles will overlap. Numerical
calculations show that for quantum numbers $n_{1}$, $n_{2}$ ranging
from 1 to 20 the direct integrals are less than one tenth of the
exchange integrals and $V_{n_{1}n_{2}},_{n_{2}n_{1}}(1,2)$ is almost
a constant, very close to $a$ (derivation around 3\% when
$0.4<a/d<1$). The bigger the values of $a/d$ is, the smaller the
ratio between the direct and the exchange integrals is.

We also calculate the interaction $V(1,2)$ for the harmonic
potential $V(r_{1,2})=r_{12}^{2}$. We find that when $a/d$ varies
from 0.7 to 0.95 the exchange integral is almost a constant with
deviations from 14\% to 4\%, and the ratio between 
the direct and exchange integrals is 0.14-0.1.

The temperature dependence of the heat conductivity $\kappa$ results
from the interplay between the specific heat and the interaction kernel.
The specific heat, defined per unit, can be easily calculated for different models, 
e.g., for a spin chain of $\omega_0$ spacing, 
$C_V=\frac{\omega_0^2}{T^2}e^{\omega_0/T}(e^{\omega_0/T}+1)^{-2}$.
In contrast, our approach does not directly bring in the temperature
dependence of the kernel $W_{\mathbf{mn}}$. Physically,
since $V_{n_{1}n_{2}},_{n_{2}n_{1}}(1,2)$ is a function of the
intersite separation, the lattice vibrations can modify it, thus
introduce temperature dependent transition rates. This effect can be
successfully included by phenomenologically introducing a
temperature dependent interaction \cite{Saxena95,Gombert02}. 
Here we adopt the simple form $V_{T}(r_{12})=V(r_{12})\gamma
(r_{12},T)$, where $T$ may be the average temperature of site 1 and
2, and $\gamma (r_{12},T)=\exp (-\alpha f(T)r_{12})$. In the case of
interacting ions \cite{Saxena95}, $\alpha $ is inversely
proportional to the Fermi velocity and $f(T)=T$. When
$f(T)\rightarrow \infty$, $V_{T}(r_{12})\rightarrow \frac{2}{\alpha
f(T)}V(0)\delta (r_{12})$. Therefore, for  high temperatures $\kappa
\propto 1/f(T)^{2}$, as the lattice specific heat typically
saturates. For the ionic solid of Ref. \cite{Saxena95} one obtains
$\kappa \propto 1/T^{2}$, in agreement with standard expectations
\cite{Peierls01}.

\textit{Summary.---} 
We have presented here a microscopic quantum derivation of
Fourier's law of heat conduction that is not limited to specifically
designed models. The derivation relays on three assumptions that are
satisfied in weak coupling and at long times for a family of exchange
interaction potentials. Our analysis naturally implies that energy
diffusion cannot emerge in harmonic models, in agreement with the 
behavior of classical systems.

\vspace{0.2in} \textbf{Acknowledgement} This work was supported by
the University of Toronto Start-up Funds.



\begin{thebibliography}{99}

\bibitem{Bonetto00} F. Bonetto, J. Lebowitz and L. Rey-Bellet,
math-ph/0002052.



\bibitem{Lepri}
S. Lepri, R. Livi, and A. Politi, Phys. Rep. {\bf 377}, 1 (2003).


\bibitem{Pereira}
E. Pereira and R. Falcao, Phys. Rev. Lett. {\bf 96}, 10061 (2006).

\bibitem{Closure}
J. Bricmont and A. Kupiainen, Phys. Rev. Lett. {\bf 98}, 214301
(2007).

\bibitem{Casati}
C. Mejia-Monasterio, T. Prosen and G. Casati, Europhys. Lett. {\bf
72}, 520 (2005).

\bibitem{Michel05} M. Michel, G. Mehler and J. Gemmer, Phys. Rev. Lett.
\textbf{95}, 180602 (2005); Phys. Rev. B {\bf 73}, 016101 (2006).

\bibitem{Michel06}
M. Michel, J. Gemmer and G. Mahler, Int. J. Mod. Phys. B {\bf 20},
4855 (2006).

\bibitem{Gemmer07}
R. Steinigeweg, H.-P. Breuer and J. Gemmer, Phys. Rev. Lett. {\bf 99}, 150601 (2007).

\bibitem{Breuer02} H.-P. Breuer and F. Petruccione, \textit{The Theory of
Open Quantum Systems} (Oxford University Press, Oxford, 2002).

\bibitem{Nakajima}
S. Nakajima, Prog. Theor. Phys. {\bf 20}, 948 (1958).

\bibitem{Zwanzig64} R. Zwanzig, Physica \textbf{30}, 1109 (1964).

\bibitem{Pereverzev06} A. Pereverzer and E. R. Bittner, J. Chem. Phys.
\textbf{125}, 104906 (2006).

\bibitem{note} V. Capek, Czech. J. Phys. \textbf{48}, 993
(1998).

\bibitem{Lin74} S. H. Lin, J. Chem. Phys. \textbf{6}, 3810 (1974).

\bibitem{GemmerPhysicaE05}
J. Gemmer, M. Michel, Physica E {\bf 29}, 136 (2005).

\bibitem{Sologubenko01} A. V. Sologubenko, K. Gianno and H. R. Ott, Phys.
Rev. B \textbf{64}, 054412 (2001).


\bibitem{Peierls01} R. E. Peierls, \textit{Quantum Theory of Solids}
(Clarendon, Oxford, 2001).

\bibitem{Noya59} H. Noya, A. Arima and H. Horie, Suppl. Prog. Theor Phys.,
\textbf{8}, 33 (1959).

\bibitem{Fetter} A. L. Fetter and John Dirk Walecka, {\it Quantum Theory of
Many-Particle Systems} (McGraw-Hill, New York, 1971).

(2004).

\bibitem{Saxena95} S. K. Saxena, R. M. Agrwal and R. P, S, Rathore, Phys.
Stat. Sol. \textbf{B 192}, 45 (1995).

\bibitem{Gombert02} M. Gombert, Phys. Rev. E \textbf{66}, 066407 (2002) and
references therein.



\end{thebibliography}
\end{document}